\documentclass[%
 reprint, 
nofootinbib,
 amsmath,amssymb,
 aps, physrev,
floatfix,
]{revtex4-2}

\usepackage{graphicx}
\usepackage{dcolumn}
\usepackage{bm}
\usepackage{xcolor}
\usepackage{verbatim}
\usepackage[normalem]{ulem} 
\usepackage{amsmath}
\usepackage[rightcaption]{sidecap}

\usepackage{url}
\usepackage{mathdots}
\usepackage{booktabs}
\usepackage{subcaption}   
\usepackage{caption}
\usepackage{siunitx}      
\DeclareMathAlphabet\mathbfcal{OMS}{cmsy}{b}{n}

\usepackage{ulem}
\usepackage[hidelinks]{hyperref}
\usepackage[autostyle, english = american]{csquotes}
\MakeOuterQuote{"}

\begin{document}

\preprint{APS/123-QED}

\title{The FRB--Galaxy Overdensity Cross-Correlation Statistic in Dispersion Space}


\author{Ryan~Raikman$^{1,2}$}
\email[]{rraikman@mit.edu}
\author{Haochen~Wang$^{1,2}$}
\author{Kiyoshi~Masui$^{1,2}$}
\author{Shion~Andrew$^{1,2}$}

\affiliation{$^1$MIT Kavli Institute for Astrophysics and Space Research, Massachusetts Institute of Technology, 77 Massachusetts Avenue Cambridge, MA 02139, USA}
\affiliation{$^2$Department of Physics, Massachusetts Institute of Technology, 77 Massachusetts Avenue Cambridge, MA 02139, USA}
\date{\today}

\begin{abstract}

Cross-correlating the dispersion of fast radio bursts (FRBs) with galaxies provides a means to study the distribution of the baryons in the Universe, even in the absence of FRB redshifts. To this end, two variants of angular cross-power spectrum statistics have been proposed: one between DM and galaxy density binned by redshift $C^{Dg}_l(z_g)$ (abbreviated $D \times g$), and one between FRB counts binned by dispersion measure (DM) and galaxy density binned by redshift $C^{fg}_l(\textrm{DM}, z_g)$ (abbreviated $f \times g$). Here we show the $D \times g$ statistic can be recovered as a DM-moment of $f \times g$, implying the latter is strictly more informative. By slicing in both DM space and galaxy redshift space, the $f\times g$ statistic separates contributions from the clustering of free electrons and from the clustering of FRB sources. We perform Fisher forecasts for FRB samples consistent with CHIME (1,600 FRBs) and the upcoming CHORD (20,000 FRBs) survey cross-correlated against the DESI Legacy Survey BGS sample and Euclid galaxy surveys, respectively. We show that, compared to the $D \times g$ statistic, the $f \times g$ statistic increases the total detection SNR by a factor of 2.5 to 4, resulting in $S/N\approx 12$ for CHIME$\times$DESI(LS) and SNR $\approx 54$ for CHORD$\times$Euclid. The $f \times g$ statistics is more sensitive to the redshift distribution of FRBs with forecasted errors on a simple parameterization of order $10 \%$. It measures the logarithmic cutoff scale for clustering of baryons due to feedback $k_{cut}$ to 26\% precision with CHIME and 14\% precision with CHORD, a slight improvement for CHIME but factor of two for CHORD. Since most FRBs currently lack host identifications, and scaling optical followup to large samples will remain challenging even with precise localizations, reliable redshifts will be unavailable for most FRBs for the foreseeable future. The $f \times g$ statistic provides a means to extract maximum cosmological information in their absence. 
\end{abstract}
\maketitle
\section{Introduction}
Fast radio bursts (FRBs) provide a unique tracer of the distribution of baryons in the Universe \cite{Lorimer_2007, Petroff_2022, Macquart_2020, Cen1999, missing_baryons, Nevalainen_2015}. These millisecond-duration extragalactic flashes acquire a frequency-dependent delay as they propagate through free electrons in diffuse plasma. This effect, integrated along the line of sight, is quantified as the dispersion measure (DM),
\begin{align} 
    \text{D}(\bm{\hat n}, z) &\equiv \int_0^{\chi(z)} \,d\chi^\prime
    \,\frac{n_e(\bm{\hat n}, \chi^\prime)}{[1+z(\chi^\prime)]^2}  \\&=\int_0^{\chi(z)} d\chi' [1+z(\chi')]n_{e, 0}[1+\delta_e(\bm{\hat n}, \chi')]
\end{align}
where $\bm{\hat{n}}$ is the observed direction, $\chi$ is the comoving distance to the source, and $n_e(\bm{\hat n}, \chi')$ is the free electron density, written in terms of the present day free electron density $n_{e, 0}$ and spatial electron overdensity $\delta_e(\bm{\hat n}, \chi')$. While contributions arise from the Milky Way, FRB hosts, and local environments, a substantial and cosmologically interesting component comes from the intergalactic medium (IGM) and gas surrounding intervening galaxies. This “cosmic” DM directly traces the bulk of the Universe’s baryons, which remain otherwise difficult to observe.

Other probes of diffuse baryons, such as the thermal and kinetic Sunyaev–Zel’dovich effects \citep{Sunyaev1972tSZ, Sunyaev1980kSZ, 2019A&A...624A..48D, guachalla2025, sunseri2025} and X-ray free–free emission \citep{pen1999, 2018Natur.558..406N} most strongly correlate to dense plasma, yielding low constraining power for diffuse clustering.  Hydrodynamic simulations further show that baryons are highly sensitive to feedback from supernovae and active galactic nuclei, processes which remain uncertain \citep{sorini_2022, Ayromlou_2023, Khrykin_sim, gebhardt2023cosmologicalbaryonspreadimpact, Velliscig_2014, elbers2025flamingoprojectcouplingbaryonic}. Measurements of the spatial distribution of baryons thus offer an opportunity to constrain both feedback models and galaxy formation physics.

Statistical approaches using FRB DMs have so far mostly involved probability distribution functions of DM over FRB redshift $z$ \cite{James_2021, flimflam, Macquart_2020,  Baptista_2024, Medlock_2024, medlock2025, sharma2025} and galaxy–FRB stacking \citep{McQuinn_2014, connor2022, Wu_2023, hussaini2025}. Recently, the first measurement of the angular cross-correlation between FRB DMs and galaxy density was made by \citet{wang2025measurementdispersionunicodex2013galaxycrosspowerspectrum}, expressing the cosmological baryon distribution in terms of the correlations of electron fields with galaxy fields \citep{masui2015, shirasaki2017, madhavacheril2019, alonso2021}. This represents a key step toward understanding the clustering of diffuse baryons via FRBs. We denote this angular cross-correlation between the DM overdensity field and galaxy overdensity field at a particular redshift $z_g$ as $C_\ell^{Dg}(z_g)$, and abbreviate as $D \times g$.

As \citet{wang2025measurementdispersionunicodex2013galaxycrosspowerspectrum}, we consider the case of an FRB population without host identifications, i.e. redshifts. Currently, most FRBs are without redshifts, and this is expected to hold in the near future. Optical follow-up, even assuming all the FRBs have outrigger-localization \cite{chime_outrigger} to the angular precision needed for confident host association, is expensive, and doing this at scale for thousands to tens of thousands FRBs will be prohibitive. 

In this work we use a generalization of $D \times g$, first presented by \citet{alonso2021} and \citet{masui2015}, and show its relation to the $D \times g$ statistic through a combination of DM moments, motivating that the generalization contains a superset of information. This statistic represents the angular cross-correlation between FRB spatial overdensity binned by DM and galaxy overdensity binned by redshift, denoted $C_\ell^{fg}(D, z_g)$, and henceforth abbreviated as $f \times g$. A drawback of the $D \times g$ statistic is that since the FRB redshifts are unknown, the inherent clustering of the FRB sources with galaxies contaminates the signal. The intuition of the generalization works around to this problem---using the measured DM as a proxy for source distance allows the degenerate contributions of FRB clustering and electron clustering to be disentangled. Unlike $D \times g$, which collapses all FRBs into a single DM field, $f \times g$ preserves information of the FRB sky position across these different DM bins. 
We further analyze its noise properties and perform Fisher forecasts, for both the CHIME \citep{CHIME2018instrument} and upcoming CHORD \citep{Vanderlinde2019_CHORD_Whitepaper} surveys, demonstrating improved constraints on model parameters describing the FRB population and clustering, relative to $D \times g$.
\section{The $f \times g$ statistic}%
\subsection{Derivation}
Here, we seek an expression for the $f \times g$ statistic in terms of the individual clustering power spectra, $P_{eg}(k)$ and $P_{fg}(k)$, and the FRB redshift distribution. We describe key parts of the derivation, and leave details and some definitions to Appendix A. We begin by seeking an expression for the overdensity of the FRB field within a certain bin of DM to cross-correlate with a galaxy field. Our underlying observable is $\tilde{n}^{(3)}_f(\bm{\hat{n}}, D)$, the FRB number density field, with each FRB labeled by a dispersion measure $D$ and direction $\bm{\hat{n}}$. We use the tilde to differentiate between fields in DM space against those in comoving distance space, with the $^{(3)}$ superscript identifies a 3-d field. 

Consider FRBs constrained to a DM bin $D$, and analogously at a comoving distance bin $\chi$. We define the expectations and overdensity fields as  
\begin{align}\label{eq:field_defn}
\tilde{n}_f^{(3)}(\bm{\hat{n}}, D) &= [1+\tilde{\delta}^{(3)}_f(\bm{\hat{n}}, D)]\bar{n}_f(D)\\
    n_f^{(3)}(\bm{\hat{n}}, \chi) &= [1+\delta^{(3)}_f(\bm{\hat{n}}, \chi)]\bar{n}_f(\chi)
\end{align}
where we have $\bar{n}_f(D)$ and $\bar{n}_f(\chi)$ as the angular number density of FRBs restricted to a bin of DM and comoving distance respectively. Considering all bins simultaneously, we have $N_f$ many observed FRBs, with the 2-d number density defined via $N_f = 4 \pi \bar{n}_f^{(2)}$.

Deriving the $f \times g$ statistic requires a translation between these two fields in Eqs. \eqref{eq:field_defn}. While $\tilde{n}_f^{(3)}(\bm{\hat{n}}, D)$ is our observable quantity, correlation functions are expressed in terms of $n_f^{(3)}(\bm{\hat{n}}, \chi)$. Suppose an FRB occurs at a comoving distance of $\chi$ in the $\hat n$ direction. In order for it to be observed in a DM bin $D$, the cosmic and host contributions to DM must equal $D$, i.e. satisfy "DM budgeting" \cite{Macquart_2020}:
\begin{align}
    \text{D} 
    &=D_{h} + \int_0^{\chi(z)} d\chi' [1+z(\chi')]n_{e, 0}[1+\delta^{(3)}_e(\hat n, \chi')]
    \\&=D_{h} + \bar{D}(\chi) + \delta D(\bm{\hat{n}}, \chi),
\end{align}

where we define the Macquart relation $\bar{D}(\chi)$ for the expected cosmic distribution, and the "Macquart perturbation" $\delta D(\bm{\hat{n}}, \chi)$  as the integral over the electron overdensity. We treat the host contribution as stochastic, and consider its redshift dependence of the probability distribution. It is usually assumed that the host contribution, denoted above as $D_h$, is redshifted (via a factor of $1/(1+z)$) as observed in our frame. However, given a potentially nontrivial time-evolution of host galaxies, we will, for now, maintain a generic redshift dependence and assume $D_h$ is drawn from probability density function $p_h(D_h|z(\chi))$. \citep{medlock2025constrainingbaryonicfeedbackeffects}. We also define the FRB distance distribution as $p_f(\chi)$ and its derivative with respect to comoving distance $p'_f(\chi)$, with the specific functional form described in Sec II. B.


We present the final result, derived fully in Appendix A, describing the angular cross-correlation $C_l^{fg}(D, z_g)$ between the FRB overdensity field in a DM bin and a galaxy overdensity field: 
\begin{widetext}

\begin{subequations}\label{eq:Clfg}
\begin{align}
     C_l^{fg}(D, z_g) &=   \bigg\{ \frac{N_f}{\bar{n}_f(D)}\int_{\chi_g}^{\infty} d\chi   \: \:p_h\left(D-\bar{D}(\chi)-\bar{D}_h(\chi)|z\right)  \frac{1}{(1+z)}\bigg( \frac{p_f(\chi)}{1+z}\frac{dz}{d\chi}- p'_f(\chi) \bigg)  (1+z_g)P_{eg}\left( \frac{l}{\chi_g}, \mu_k=0, z_g \right)\frac{1}{\chi_g^2} \bigg \} \label{eq:Clfg_background}\\ &\quad
    + \bigg \{\frac{N_f}{\bar{n}_f(D)}  \: p_f(\chi_g)\:p_h\left(D-\bar{D}(\chi_g)-\bar{D}_h(\chi_g)\right|z_g)  \left(P_{fg}\left(\frac{l}{\chi_g}, \mu_k=0, z_g\right) - P_{eg}\left( \frac{l}{\chi_g}, \mu_k=0, z_g\right) \right)\frac{1}{\chi_g^2} \bigg\} .\label{eq:Clfg_contact}
\end{align}
\end{subequations}
\end{widetext}

To visualize the different contributions to the signal spectrum, we individually inspect the two terms in the $f\times g$ cross-correlation, the "background", Eq. \eqref{eq:Clfg_background} and "contact", Eq. \eqref{eq:Clfg_contact} terms respectively. The background term considers FRBs passing through the specified galaxy redshift, and interacting with the electrons locally correlated with those galaxies. As such, we expect non-negligible contributions only if the DM bin is at an effective redshift greater than that of the galaxies. The contact term considers interactions between the spatial distribution of galaxies and FRB progenitors, which we expect to be non-negligible only if the DM bin is at the same effective redshift of the galaxies. We also work under the assumption that the Milky Way contribution to the DM has been sufficiently subtracted, and expect that errors introduced by this subtraction do not correlate with large-scale structure.
\begin{figure*}[ht]
    \centering
    \includegraphics[width=0.7\textwidth]{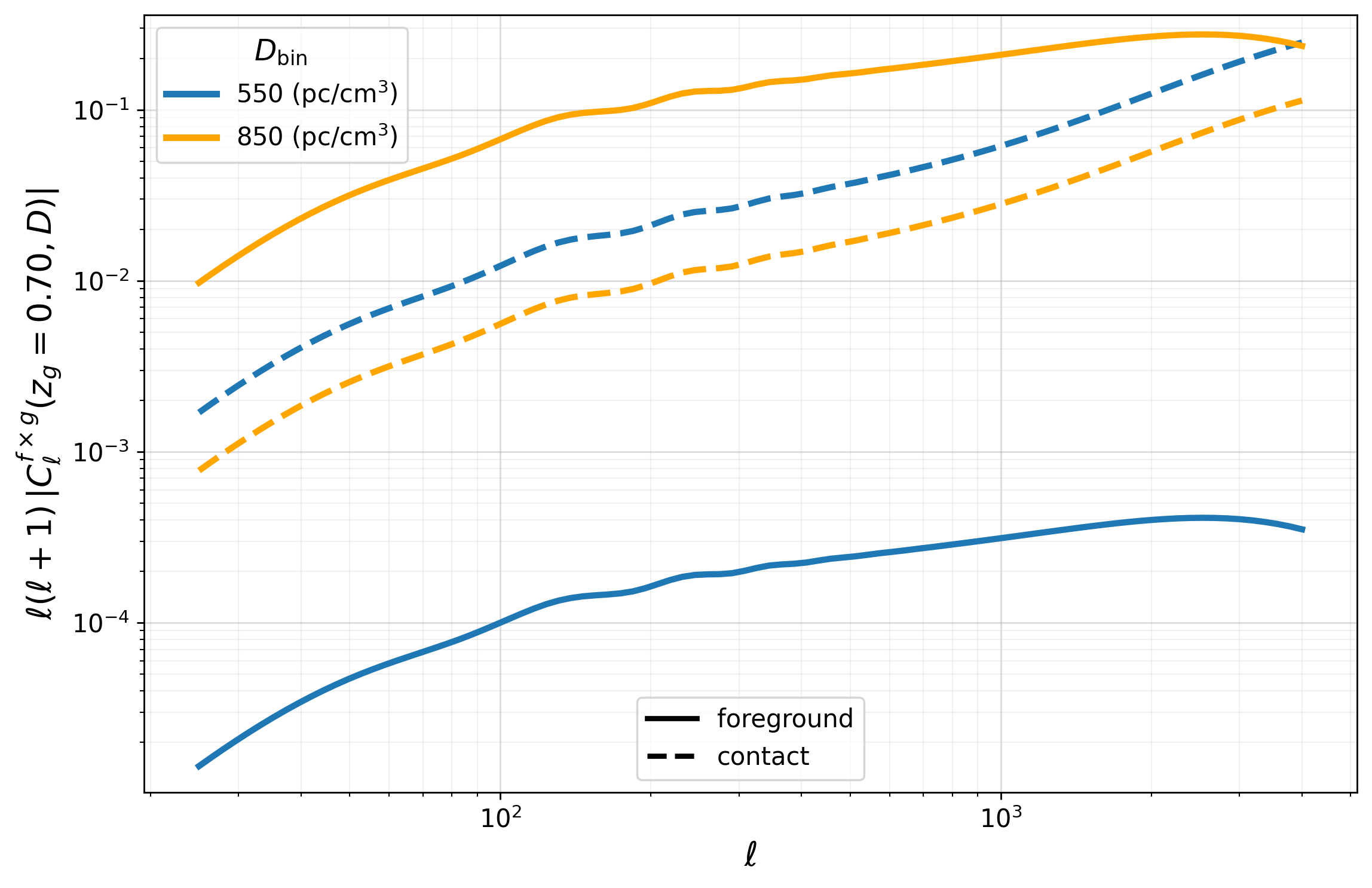}
    \caption{
    $f\times g$ angular cross-power spectrum for a galaxy overdensity field centered at redshift $z_g = 0.7$ against an FRB overdensity field at two different DM bins. The two choices of DM bin represent FRBs approximately at and behind the galaxy redshift, as extrapolated by the Macquart relation. Solid lines show the "background", Eq. \eqref{eq:Clfg_background} contribution from cross-correlations of the galaxy field with ionized electrons as measured by the DM. Dashed lines show the "contact" term, Eq. \eqref{eq:Clfg_contact}, arising from the inherent clustering of FRB progenitor positions with those of galaxies, expected since FRB progenitors tend to reside in or near galaxies. Not shown in this plot is the cross correlation against $D_{\text{bin}}=350 \; \text{pc}/\text{cm}^3$, representing FRBs in front of galaxies. As expected, this does not contribute to the cross correlation, i.e. $C_{\ell}^{fg}(z_g=0.7, D=350)=0$. }   
    \label{fig:Cl_fDxg}
\end{figure*}

\subsection{Power Spectra and FRB Population Model}
Computing spectra of the $f \times g$ statistic in Eq. \eqref{eq:Clfg} requires models of the FRB population with redshift $p_f(\chi)$, clustering power spectra $P_{eg}(k, z)$ and $P_{fg}(k, z)$, and host contribution distribution, $p_h(D|z)$. We use parameterizations identical to \citet{wang2025measurementdispersionunicodex2013galaxycrosspowerspectrum}: a Schechter luminosity function for the observed FRB population, and power spectra described via an electron bias $b_e$, FRB bias $b_f$, and exponential scale cutoff $k_{\text{cut}}$. Respectively, these correspond to Eq. (46) and Eqs (6), (7) used by \citet{wang2025measurementdispersionunicodex2013galaxycrosspowerspectrum}. We also choose to include $b_e$ as a free parameter, even though it is expected to be close to 1 \citep{Shaw_2010}. Our Fisher forecasts show that it is not strongly degenerate with other parameters, so including $b_e$ does not degrade the forecast. Additionally, its forecasted uncertainty is a useful proxy for how well the "background" and "contact" terms are distinguished. In practice, a tight prior would be enforced. 

For the FRB redshift distribution, we have
\begin{align}
    dn(L) &= \phi^* \left( \frac{L}{L_*} \right)^{\alpha} e^{-L / L_*} \, d\left( \frac{L}{L_*} \right)\\
    L_* &= F_{\mathrm{th}} \, 4\pi d_L^2(z_*),
\end{align}
where $\phi_*$ and $F_{\mathrm{th}}$ are normalization factors, and $\alpha, z_{*}$ are model parameters, representing the power law index and characteristic redshift respectively. This model describes a power-law distribution with an exponential cutoff towards high $z$ and peaking around $z_{*}$. While the true underlying distribution is likely more complicated, this serves as a simple ad hoc model for this work, sufficient to capture the different information content between the $f \times g$ and $D \times g$ statistics.
For the clustering power spectra we have, 
\begin{align}
    P_{eg}(k, z) &= b_eb_g(z) e^{-k/k_{\text{cut}}} P(k, z)\\
    P_{fg}(k, z) &= b_fb_g(z)P(k, z),
\end{align}
where $P(k, z)$ is the non-linear halo-fit matter power spectrum \citep{Takahashi_2012}, and parameters $b_e, b_f, k_{\text{cut}}$ are to be fit. We take $b_g(z)$ as known from galaxy-galaxy clustering measurements, and simplify by using a fiducial $b_g=1.2$ across all redshifts. We use a value of $b_f=2.4$ and $k_{\text{cut}}=1.6\;  h/\text{Mpc}$, coming from the maximum of the posterior produced by \citet{wang2025measurementdispersionunicodex2013galaxycrosspowerspectrum}. As the "contact" term Eq. \eqref{eq:Clfg_contact} has a relatively small overall contribution, the forecasts are not very sensitive to $b_f$. These power spectra models qualitatively describe the expected behavior---FRB progenitors follow the overall matter distribution with a large scale bias $b_f$, and electrons cluster around galaxies up to some cutoff scale $k_{\text{cut}}$, where feedback processes destroy small-scale power. Given the uncertainty of feedback processes in galaxies, this simple choice of parametrization serves as a first step in discriminating between different feedback prescriptions. 
For the host contribution, we take a log-normal distribution with mean and variance $\mu_{\text{host}}, \sigma^2_{\text{host}}$, and implement a simple redshift scaling,
\begin{align}
p_{\text{host}}(D, z) &=
\frac{1}{D \, \bar{\sigma}(z) \sqrt{2\pi}}
\exp\!\left[-\frac{(\ln D - \bar{\mu}(z))^2}{2\bar{\sigma}(z)^2}\right] \\
\bar{\mu}(z) &= \frac{1}{1+z}\ln\!\left(\frac{\mu_{\text{host}}^2}{\sqrt{\sigma_{\text{host}}^2 + \mu_{\text{host}}^2}}\right) \notag \\
\bar{\sigma}(z)^2&= \frac{1}{(1+z)^2}\ln\!\left(1 + \frac{\sigma_{\text{host}}^2}{\mu_{\text{host}}^2}\right), \notag
\end{align}
with parameters $\mu_{\text{host}}$, $\sigma_{\text{host}}^2$, representing the $z=0$ mean and variance of the host DM distribution, to be fit. For this work, we choose fiducial values of $\mu_{\text{host}}=100$ and $\sigma_{\text{host}}=100$ pc/$\text{cm}^3$ \citep{Bernales_Cortes_2025, Reischke_2025}.

\subsection{Fiducial Parameters}
To visualize the theoretical signal template, we require a set of fiducial parameters. For the FRB redshift distribution parameter $\alpha$, we use the maximum of the posterior distribution computed by \citet{wang2025measurementdispersionunicodex2013galaxycrosspowerspectrum}. For $z_*$, we take a value that roughly agrees with the CHIME catalog 2 observed DM distribution, with log-normal distributions for the cosmic and host contributions, following the methodology of \citep{Shin_2023}. Parameters $\alpha, z_{\star}$ for the CHORD survey are derived from an MCMC simulating detection efficiency of FRBs with  instrumental effects taken into account. While we do not expect the true FRB redshift distribution to follow such a simple form, this serves as a starting point in characterizing our capacity to measure cosmological and astrophysical parameters of interest. We summarize these choices in Table \ref{tab:fid_params}:
\begin{table}[h]
  \centering
  \renewcommand{\arraystretch}{1.2}
  \begin{tabular}{lccccc}
    \hline
    Instrument & $z_\star$ & $\alpha$ & $N_{FRB}$ & $N_{gal}$ & $f_{\text{sky}}$ \\
    \hline
    CHIME x DESI & 0.6 & 0.1 & 1{,}630 & 5.8$\times 10^6$ & 0.275 \\
    CHORD x Euclid& 2.15 & -1.5 & 20{,}000 & 50$\times 10^6$ & 0.3\\
    \hline
  \end{tabular}
  \caption{FRB survey parameters. CHIME parameters come from \citet{wang2025measurementdispersionunicodex2013galaxycrosspowerspectrum} and fitting the Catalog 2 DM distribution. CHORD parameters come from a MCMC simulation of detector efficiency and instrumental effects.} \label{tab:fid_params}
\end{table}

For the DESI galaxy redshift distribution, we take the BGS population from the photometric legacy survey \citep{Dey_2019}. We use 5.8 million galaxies and an overlap region of $f_{\text{sky}}=0.275$, computed by explicitly overlaying the galaxy and FRB fields. For the Euclid fiducial galaxy redshift distribution, we take a simple model with parameter choices used by \citet{Kirk_2013}:
\begin{align}
    n(z) \propto z^{\alpha_g} \exp \left(-\frac{z}{z_0}\right)^{\beta_g},
\end{align}
taking 50 million galaxies in an overlapping region of $f_{\text{sky}}=0.3$ following a rough calculation of the slightly increased sky coverage of both the CHORD and Euclid surveys. Using these parameterizations, we visualize our theoretical signal template. In Figure 1 we show the different components of $C_{\ell}^{fg}(D, z_g)$, Eq. \eqref{eq:Clfg}, for specific choices of DM and $z_g$ bins, using fiducial parameters for CHIMExDESI as presented in Table \ref{tab:fid_params}.

\subsection{Reduction to $C_l^{Dg}(z_g)$ Statistic}
Here, we present the connection between the $f \times g$ and $D \times g$ statistics. Since both use the same underlying DM field, the distinction between the two statistics lies in their construction. Whereas $D \times g$ uses the entire DM field at once, $f \times g$ splits the DM field into many bins. By doing so, $f \times g$ individually resolves specific terms of the cross-correlation, as demonstrated in Fig. \eqref{fig:Cl_fDxg}. Nonetheless, by combining $f \times g$ across these DM bins in a particular fashion, we can exactly recover the $D \times g$ statistic constructed from the entire DM field. We now demonstrate this connection, recovering $C_l^{Dg}(z_g)$ as a linear combination of DM-weighted integrals over $C_l^{fg}(D, z_g)$, henceforth referred to as "moments". As such, we prove that the $f \times g$ statistic must contain more information than $D \times g$. We define the $n$th DM-weighted moment of $f \times g$ as
\begin{align}
    M[D^n \:C^{fg}_l(D)] &\equiv \int D^ndD \:C^{fg}_l(D) \frac{n_f(D)}{N_f}.
\end{align}
We also define the theoretical average observed DM across the entire FRB population:
\begin{align}\label{Df}
    \bar{D}_f &= \int d\chi p_f(\chi) [\bar{D}(\chi) + \bar{D}_h(\chi)].
\end{align}
We expect that the correct combination of DM-weighted moments should represent the construction of the DM field used in $D \times g$ from the FRB field binned by DM as used in $f \times g$. Namely, we weight by a single power of DM, and subtract off the mean DM value, $\bar{D}_f$. With explicit computations of the relevant $n=0, n=1$ moments in Appendix B, we demonstrate the match to Eq. 6 found by \citet{wang2025measurementdispersionunicodex2013galaxycrosspowerspectrum}.

\begin{align}\label{dxg}
    & M\left[D^1\:C_\ell^{fg}(D, z_g)\right] - \bar{D}_f\:M\left[D^0 \:C_\ell^{fg}(D, z_g) \right]=\notag \\
    &\qquad n_{e,0}(1 + z_g) \, P_{eg} \left( \frac{\ell}{\chi_g}, z_g \right) \, \frac{1}{\chi_g^2}
\int_{\chi_g}^{\infty} d\chi\, p_f(\chi) \notag\\
&\qquad + \: p_f(\chi_g) \left( \bar{D}(\chi_g) + \bar{D}_h(\chi_g) - \bar{D}_f\right)
P_{fg} \left( \frac{\ell}{\chi_g}, z_g \right)\frac{1}{\chi_g^2}\notag \\
& \qquad = C_l^{Dg}(z_g)
\end{align}

\section{Forecasts}
In this section we outline the framework and calculations necessary to forecast the constraining power of the $f \times g$ and $D \times g$ statistics. We begin by computing the noise spectra of both statistics, followed by a description of the Fisher formalism used to compute the parameter uncertainties for both the CHIME and CHORD surveys.
\subsection{Power Spectrum Uncertainties}

To compute the noise spectrum, we start with the four-point function corresponding to the noise of the $f \times g$ statistic. First, given the large inherent variance in the DM field, we assume that the four-point function can be decomposed into the disconnected pieces to good accuracy. We have checked this approximation, and the disconnected pieces even dominate the $+2C_{\ell}^{fg}(D, z_g)$ cross-term.
\begin{align}
    (N_{\ell}^{fg}(D, z_g))^2 &= \langle a_{lm}^g a_{lm}^{g*}  a_{lm}^f(D) a_{lm}^{f*}(D') \rangle \\
    &\approx \langle a_{lm}^g a_{lm}^{g*} \rangle \langle a_{lm}^f(D) a_{lm}^{f*}(D') \rangle
\end{align}
These expressions, under the assumptions of Gaussian errors, enable forecasting both the error-bars (via Eq 6 from \citet{madhavacheril2019}) on cross-correlation measurements and uncertainties of model parameters. For the galaxy field, we have \citep{madhavacheril2019}:
\begin{equation}\label{clgg}
    \langle a_{lm}^g a_{lm}^{g*} \rangle_{\text{bin}} = \frac{1}{\chi_g^2 (\Delta \chi_g)} P_{gg}(k, z_g) \big|_{k = \ell / \chi_g} + \frac{1}{n_g^{\mathrm{2d}}},
\end{equation}
where we define $\Delta \chi_g$ as the width of the galaxy distribution. Given a large galaxy population, both $P_{gg}$ and shot-noise terms are relevant. In contrast, the intrinsic noise of the FRB field binned in DM is dominated by the shot-noise and line of sight variance terms for the fiducial number of sources used in this forecast (verified by taking explicit autocorrelation of $\delta_f$ terms). Hence, the autocorrelation of $\delta_f(D)$ takes a simple expression, cleanly diagonalized once averaged over two DM bins $D_b, D_b'$:
\begin{align}\label{clfDfD}
    &\langle a_{lm}^f(D) a_{lm}^{f*}(D') \rangle \approx \frac{\delta_D(D - D')}{n_f(D)}\notag \\
    &\langle a_{lm}^f(D_b) a_{lm}^{f*}(D_b') \rangle_{\text{bin}}=  \frac{1}{\Delta D}  \frac{\delta_{D_b', D_b}}{n_f(D_b)}.
\end{align} 
Combining these two expressions, we have the $f \times g$ noise spectrum as
\begin{align}
     (\mathcal{N}_{l}^{fg}(D, z_g))^2 &\approx \left(\frac{P_{gg}\left(k = \frac{\ell}{\chi_g}, z_g\right)}{\chi_g^2 (\Delta \chi_g)}  + \frac{1}{n_g^{\mathrm{2d}}}\right)\frac{1}{\Delta D}  \frac{\delta_{D_b', D_b}}{n_f(D_b)}.
\end{align}
We then use this to constrain the SNR across a bin of $\ell$ modes, via
\begin{equation}
    \mathrm{SNR}_{\mathrm{bin}}(D_b, z_g)^2 = \Omega \int_{\ell_{\min}}^{\ell_{\max}} \frac{\ell \, d\ell}{2\pi}  \left(\frac{C_\ell^{fg}(D_b, z_g)}{\mathcal{N}_\ell^{fg}(D_b, z_g)} \right)^2.
\end{equation}

\begin{figure*}[ht]
    \centering
    \includegraphics[width=0.7\textwidth]{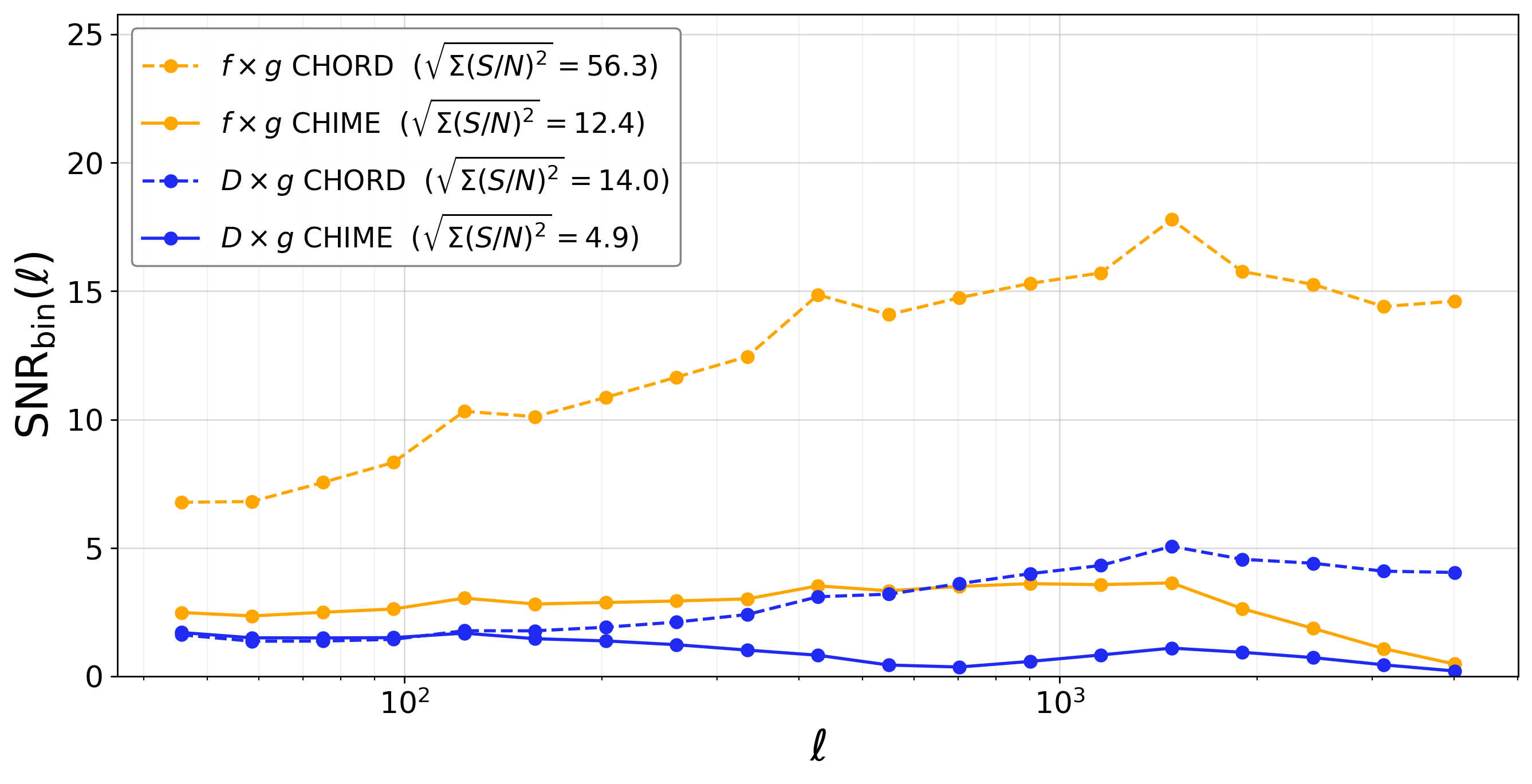}
    \caption{Comparison of overall SNR present across $\ell$ bins for both the $D \times g$ and $f\times g$ statistics, for both the CHIME and CHORD surveys. We see that the $f \times g$ contains higher SNR across both surveys as compared to $D \times g$, consistent with $f \times g$ containing more information as a superset of $D \times g$. }
    \label{fig:collapsed_comparison}
\end{figure*}

It remains to compute the noise spectrum of the DM overdensity field for the $D \times g$ statistic. Analogous to the procedure to map $C_l^{fg}(D, z_g)$ into $C_l^{Dg}(z_g)$, we take the same linear combination of moments as Equation 21. We first define $R(D)$ as used to collapse the $f \times g$ into the $D \times g$ statistic,
\begin{align}
R(D) &\equiv (D-\bar{D}_f) \frac{\bar{n}_f(D)}{N_f} \notag
\\  C_l^{Dg}(z_g) &= \int dD\: \: R(D) \: C_l^{fg}(D, z_g)  \notag,
\end{align}
where the explicit form for $\bar{n}_f(D)$ in Eq. \eqref{eq:nfD} is useful. We then apply this same scheme to the noise spectrum: 
\begin{align}\label{clDD}
    \langle a_{lm}^D a_{lm}^{D*}\rangle &= \bigg[ \int dD \: R(D) \int dD' R(D') \bigg] \frac{\delta_D(D - D')}{{n}_f(D)}\notag\\
 &=\frac{1}{\bar{n}^{2d}_f}\int d\chi p_f(\chi) \int dD \: D^2 \: p_h(D|z(\chi) )\notag \\ & \quad+ \:\frac{1}{\bar{n}^{2d}_f}\int d\chi p_f(\chi) \left[  \bar{D}(\chi) + \bar{D}_h(\chi) - \bar{D}_f\right]^2 \notag \\
 &\equiv \frac{1}{\bar{n}^{2d}_f}\sigma_{\text{host, effective}}^2 + \frac{1}{\bar{n}^{2d}_f}\sigma^2_{z}.
\end{align}
Intuitively, the first term represents the contribution to the variance of the DM field from the intrinsic variance of the host distribution, considering its evolution over redshift. The second term, introduced by \citet{Reischke_2025}, can be understood as the uncertainty of replacing each FRB's true distance with an expectation over the parametrized FRB redshift distribution. As a sanity check, treating the FRBs as coming from a thin shell (i.e. replacing $p_f(\chi) \rightarrow \delta(\chi-\chi_f)$) drops this term, reducing the expression to $\frac{1}{\bar{n_f^{2d}}} \sigma_{\text{host}}^2$, in agreement with \citet{madhavacheril2019}. In Figure \ref{fig:collapsed_comparison}, we forecast the SNR for different $\ell$ bins as optimally weighted across $z_g$ bins ($z_g$ and $D$ bins) for the $D \times g$ ($f \times g$) statistics, for both the CHIME and CHORD surveys. To quantify the effect from localization uncertainty of FRBs in CHIME, we match \citet{wang2025measurementdispersionunicodex2013galaxycrosspowerspectrum}, and take $C^{Dg}(\ell) = C_{\text{theory}}^{Dg}(\ell) \exp[-l^2/(2 \ell_{\text{loc}}^2)]$, adopting $\ell_{\text{loc}}=2000$. We generally consider the $\ell$ range from $\ell_{\text{min}}=40$ to $\ell_{\text{max}}=6000$. The lower bound of $\ell=40$ is chosen to bypass effects of survey geometry and the Milky Way DM contribution. The upper bound is irrelevant to the CHIME survey given the localization error, and is taken to very roughly account for the localization errors in CHORD. The $f \times g$ cross-correlation picks up an identical suppression by $\ell_{\text{loc}}$ for the CHIME case. In our Fisher forecasts, we include this suppression, but unlike \citep{wang2025measurementdispersionunicodex2013galaxycrosspowerspectrum}, we take this value as given rather than marginalizing over its value. 
\subsection{Fisher Formalism}
Here, we are interested in forecasting the constraining power of the two statistics, $f \times g$ and $D \times g$, on model parameters, and how that constraining power compares. To do so, we first construct the full covariance matrix for each statistic, using both the cross-correlation/signal model (Eq. \eqref{eq:Clfg} for $f \times g$ and Eq. \eqref{dxg} for $D \times g$) and auto-correlation/noise models (Eq. \eqref{clfDfD} for $f \times g$, Eq. \eqref{clDD} for $D \times g$, and Eq. \eqref{clgg} for both). Explicitly, we have
\begin{align}
\text{Cov}^{f\times g}(\ell)
&=
\begin{bmatrix}
\{C^{ff}_{\ell}(D_{b, i}\:,\: D_{b, j})\} & \{C_{\ell}^{fg}(D_{b, i} \:,\: z_{g, m}) \} \\
\{C_{\ell}^{fg}(D_{b, j} \:,\: z_{g, n})\} & \{C_{\ell}^{gg}(z_{g, m} \:,\:z_{g, n})\}
\end{bmatrix}\\
\text{Cov}^{D\times g}(\ell)
&=
\begin{bmatrix}
C^{DD}_{\ell}  & \{C_{\ell}^{Dg}( z_{g, m})\} \\
\{ C_{\ell}^{Dg}(z_{g, n})\} & \{C_{\ell}^{gg}(z_{g, m} \:,\:z_{g, n})\}
\end{bmatrix}
\end{align}
Here, $\text{Cov}(\ell)$ is a block diagonal matrix as a function of angular multipole $\ell$, with $\{ \cdots \}$ explicitly indicating a matrix quantity. DM bins $D_b$ are indexed with letters $i, j$ each in $[1...d]$ where $d$ is the number of DM bins, and galaxy redshift bins $z_g$ are indexed with $m, n$ each in $[1...G]$ where $G$ is the number of galaxy redshift bins. Together, $\text{Cov}^{f\times g}(\ell)$ is a $(d+G) \times (d+G)$ matrix, and $\text{Cov}^{D\times g}(\ell)$ is a $(1+G)\times(1+G)$ matrix. In this form, we have explicitly assumed Gaussian error for the quantities of interest (as opposed to, say, estimating them from multiple simulation mocks), and therefore they represent lower bounds to the true noise. 

Next, we construct the Fisher matrix, which corresponds to profiling curvature of $\chi^2(\lambda_i)$, i.e. the goodness of fit, across the values of our model parameters $\lambda_i$. When the value of some parameter $\lambda_i$ is perturbed, if $\chi^2(\lambda_i)$ changes rapidly, we are "sensitive" to its value, and therefore forecast a low error for $\lambda_i$, and vice versa. Explicitly,

\begin{equation}
    F_{ab} =f_{\text{sky}}  \int dl \frac{2l+1}{2}\: \mathrm{Tr} \left[ \text{Cov}(l)_{,a} \text{Cov}(l)^{-1} \text{Cov}(l)_{,b} \text{Cov}(l)^{-1} \right],
\end{equation}
where $f_{\text{sky}}$ is the fractional sky coverage of the overlapping region of the FRB and galaxy fields to be cross-correlated. Superscript $^{-1}$ indicates a matrix inverse, and subscript $_{,a}$ denotes a derivative with respect to parameter $a$. 

On taking derivatives, we take care only to take the derivatives of quantities that (in data analysis) are directly model dependent. For example, with the case of $D \times g$, to get a value for the average observed DM, we use Eq. \eqref{Df}, which involves the FRB redshift distribution parameters. However, in practice, this value is fixed by the observed DM field, i.e. we do not fit the FRB redshift distribution parameters to its value. In the case of $f \times g$, the same goes for the values of $\bar{n}_f(D_b)$. In principle, this quantity across DM bins also encodes a lot of information about the FRB redshift distribution, but as it is fixed by the measurement and potentially more susceptible to DM-dependent selection effects, we do not take its derivative in the Fisher forecast. 

\subsection{Parameter Forecasts}
\begin{figure*}[t]
  \centering
  \includegraphics[width=\linewidth,
    trim=0pt 0pt 0pt 0pt,clip]{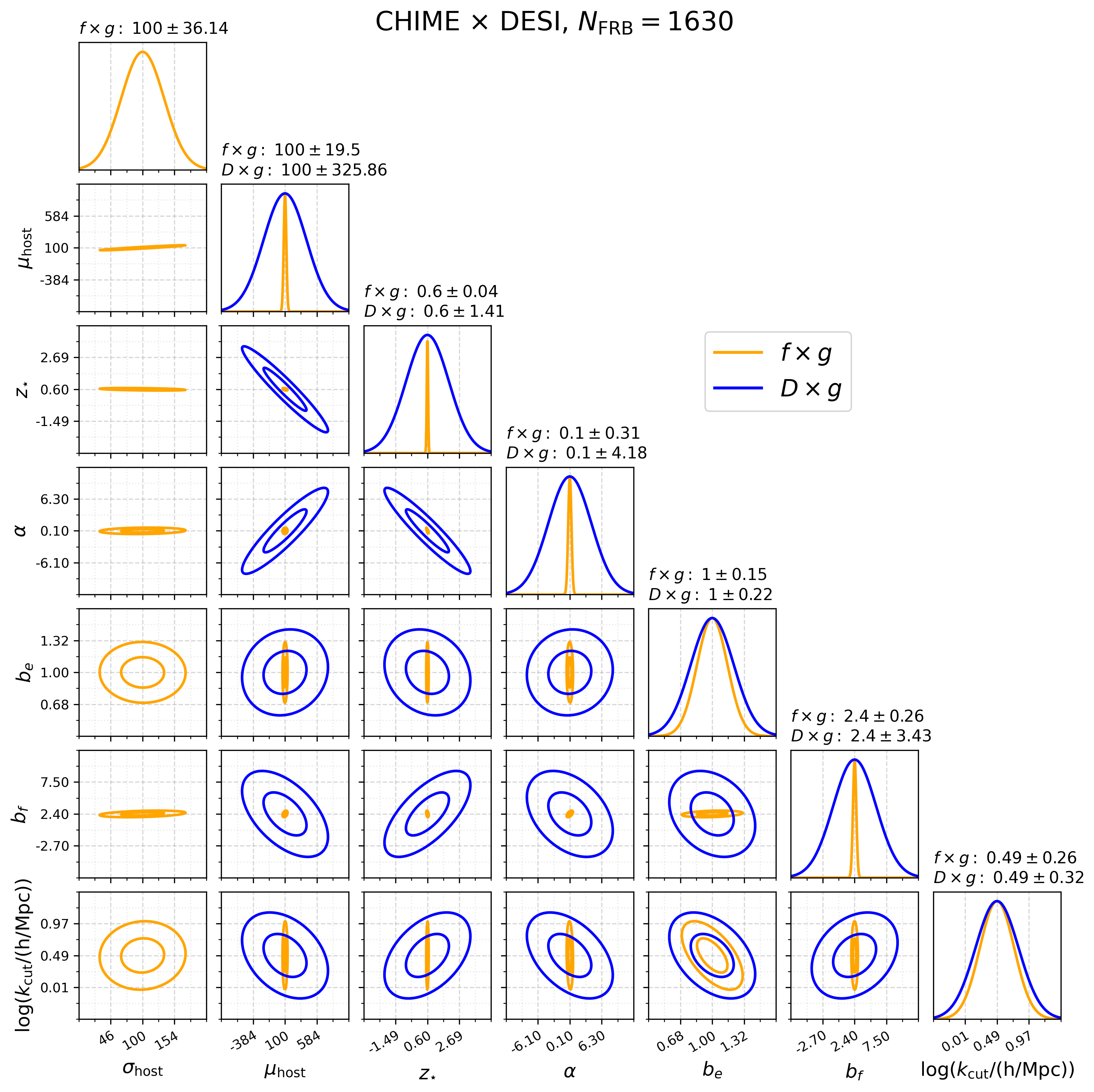}
  \caption{Comparison of forecast posteriors describing the host DM distribution, FRB redshift distribution, and $P_{eg}$, $P_{fg}$ clustering power spectra from the $D\times g$ and $f\times g$ statistics for the CHIME  survey. $f \times g$ presents a tighter posterior distribution as compared to $D \times g$ for every parameter, as expected. $D \times g$ has no dependence on $\sigma_{\text{host}}$, hence the lack of a blue posterior in the left-most column. }
  \label{fig:CHIME_posterior}
\end{figure*}

\begin{figure*}[t]
  \centering
  \includegraphics[width=\linewidth]{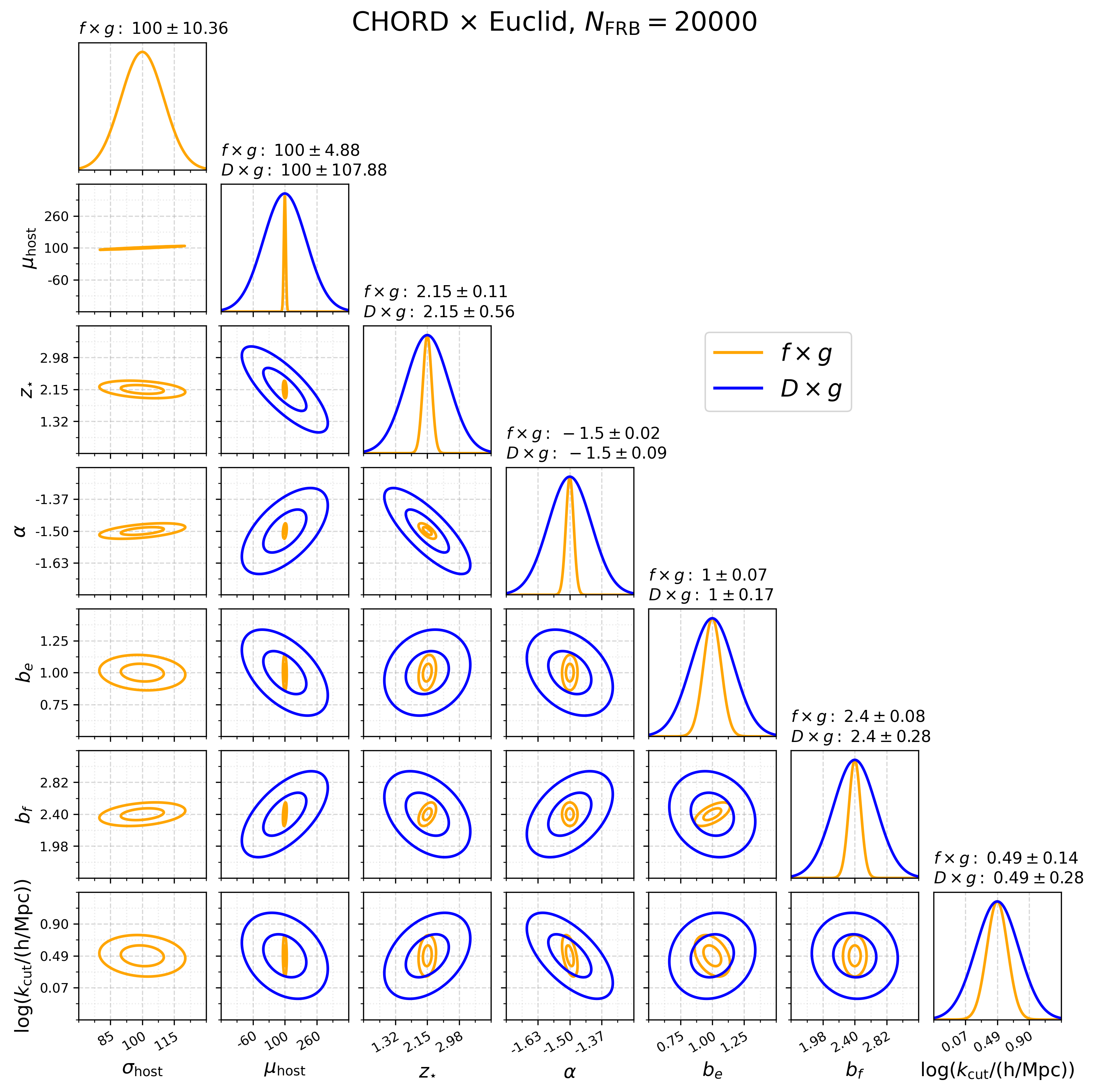}
  \caption{Same as Fig.~\ref{fig:CHIME_posterior}, but for the CHORD survey. Going from CHIME to CHORD populations presents an increase in constraining power for every parameter.}
  \label{fig:CHORD_posterior}
\end{figure*}

We present the comparison between estimated uncertainties for FRB distribution and clustering parameters in a corner plot for the CHIME$\times$DESI in Figure \ref{fig:CHIME_posterior} and for the CHORD$\times$Euclid surveys in Figure \ref{fig:CHORD_posterior}.

\section{Discussion}
Our Fisher forecast shows an improvement on the constraining power of the $f \times g$ statistic relative to the $D \times g$ statistic for model parameters describing the redshift distribution of FRB sources as well as clustering statistics of electrons with galaxies. This gain comes from using DM as a proxy for distance, which enables the "background" and "clustering" terms to be individually resolved. The first consequence is a tighter bound on $b_f$ from the $f \times g$ statistic, which in turn breaks degeneracies with $b_f$ present in the $D \times g$ statistic. Additionally, as the strength of cross-correlation between a DM bin and galaxy redshift bin directly depends the number of FRBs behind those galaxies, $f \times g$ is more sensitive to the FRB redshift distribution parameters $\alpha$ and $z_{\star}$. We also forecast that $f \times g$ enables a strong improvement in measuring $\mu_{\text{host}}$. This comes from the fact that $f \times g$ is sensitive to a change in $\mu_\text{host}$ at every DM bin, whereas $D \times g$ only indirectly probes $\mu_{\text{host}}$ through its rescaling of the $P_{fg}$ term. Finally, $f \times g$ allows for a constraint on $\sigma_{\text{host}}$, since the scattering of FRBs into different DM bins is measured, whereas $D \times g$ has no dependence on $\sigma_{\text{host}}$.


However, some caveats remain in the transition from the $D \times g$ to $f \times g$ statistic. The DM binning scheme introduces a potential source of information loss, since the relative DM fluctuations within a bin are neglected. However, given that the DM from a single source already contains intrinsic uncertainty from variance in the host contribution, we expect that choosing DM bins of order the width of the host contribution should preserve most information, a prediction which can be tested by mocking this analysis with a simulation.  Given a hypothetical range of DM of 0 to about 2000 DM pc $\text{cm}^{-3}$, \citep{chime_catalog_2021, chime_cat_2}, and a host width of 100 DM units \citep{Bernales_Cortes_2025, Reischke_2025}, this then translates an increase in computation by a factor of 20 relative to the $D \times g$ statistic.

Both $f \times g$ and $D \times g$ are potentially susceptible to instrument-dependent DM selection effects, yielding an observed DM distribution inconsistent with the true astrophysical one \citep{cheng2025exploringselectionbiasesfrb}. While it is not investigated in this paper, theoretically more information exists in the high-DM sources, since they correlate with more foreground galaxy bins. Relatively losing these FRBs, as compared to low-DM FRBs, could therefore reduce the capacity to constrain model parameters. Given that the $f \times g$ statistic relies on the binning by DM, it could potentially be more susceptible to these effects. The impact of such selection effects can be investigated through simulations to understand how they effect the constraints on model parameters in a similar fashion as \citet{cheng2025exploringselectionbiasesfrb}. 

The relatively simple model parametrizations of both the FRB spatial distribution and power spectra are also likely insufficient to capture the underlying reality. The tiny error bars reported in the Fisher forecast for $f \times g$ on the $\alpha$ and $z_{\star}$ parameters are likely artifacts of this very simple model. However, with a larger source population, or by localizing the FRBs, the redshift distribution model can be more precisely constrained and refined. 

Additionally, expanding the electron and galaxy fields with effective field theories would yield a physically motivated and consistent parametrization of the non-linear power spectra. Our simple and ad-hoc $k_{\text{cut}}$ parameterization is not physically motivated, serving as a starting point for these kinds of analyses. As FRB samples grow, this work can also be improved by moving beyond the non-linear halofit model in modeling the $f\times g$ cross-correlation, for example by directly incorporating one–loop perturbation theory \citep{cabass2022snowmasswhitepapereffective, ivanov2022effectivefieldtheorylarge} or a hybrid treatment as by \citet{ Sullivan_2021}, allowing a more accurate model of the weakly non-linear regime where constraining power resides.
\section{Conclusion}
In this work we have developed the $f \times g$ statistic, describing the cross-correlation of the FRB overdensity field binned by dispersion measure against a galaxy overdensity field in a redshift bin. Via a Fisher forecast for FRB populations consistent with CHIME and upcoming CHORD surveys, we have shown that the $f \times g$ statistic yields smaller predicted uncertainties on all model parameters as relative to the $D \times g$ statistic. These parameters describe the FRB progenitor distribution with redshift, the statistical clustering of electrons and galaxies $P_{eg}(k, z)$, and the clustering of FRB progenitors and galaxies $P_{fg}(k, z)$. Additionally, the $f \times g$ statistic constrains the host DM distribution more tightly through its direct knowledge of the DM field itself. 
In contrast to tSZ/kSZ or free-free emission analyses, analyses of the DM field such as $f \times g$ provide a unique observational probe of the baryon density field itself. Results from a DM field analysis can serve as priors to SZ/X-ray analyses to break degeneracies in the relevant weighting fields, for example as by \citet{madhavacheril2019} to constrain large-scale momentum fields when combined with kSZ measurements. Additionally, constraints on the form of the $P_{eg}(k, z)$ can be directly compared to simulations with various feedback mechanisms, helping differentiate feasible mechanics from infeasible ones.
This kind of analysis could also provide information on the nature of FRB progenitors. Measurements of their redshift distribution, here parametrized by $z_{*}$ and $\alpha$, could indicate whether progenitors trace star formation or potentially are sourced by a completely different mechanism. Measuring the galaxy-FRB clustering $P_{fg}(k, z)$, particularly on large scales, could indicate if progenitors tend to reside within specific kinds of galaxies (high or low bias). Finally, the $f \times g$ statistic yields insight into the statistics of the contribution to DM from the host, shedding insight on the nature of the FRB host and its environment. 
While the aforementioned improvements to modeling can be made, this work demonstrates the increase in constraining power going from the $D \times g$ to $f \times g$ statistic. In turn, this adds to understanding of large-scale structure through setting priors for other probes, in galactic astrophysics by directly measuring feedback mechanisms, and in FRB astronomy by better understanding the progenitor redshift distribution and host environments. 
\begin{acknowledgments}
We thank Seth Siegel for providing the code used to simulate the FRB
population observed by the CHORD instrument. We also thank Jamie Sullivan
for useful discussions. This work is supported by NSF grant 2008031.
\end{acknowledgments}
\appendix
\onecolumngrid
\section{$C_l^{fg}(D, z_g)$ Derivation}
    Here we detail the derivation of the angular cross-correlation of FRB spatial overdensity at a particular DM bin against galaxy overdensity at a particular redshift shell. We begin by formalizing the observation that the spatial FRB density field $\delta^{(3)}_f(\bm{\hat{n}}, \chi)$ is "scattered" into the DM FRB density field $\tilde{\delta}^{(3)}_f(\bm{\hat{n}}, D)$ according to the "DM budgeting" principle, Eq 6:

    \begin{align}
        \tilde{n}_f^{(3)}(\bm{\hat{n}}, D) = \int d\chi \: \delta^{D}(D-D_h -\bar{D}(\chi)-\delta D(\bm{\hat{n}}, \chi)) \:  p_f(\chi) \: n^{(3)}_f(\bm{\hat{n}}, \chi).
    \end{align}
    Since $D_h$ is a random variable, we replace the Dirac delta function with the host DM probability distribution:
    \begin{align}
        \tilde{n}_f^{(3)}(\bm{\hat{n}}, D) = \int d\chi \: p_h \bigg(D -\bar{D}(\chi)-\delta D(\bm{\hat{n}}, \chi)| z(\chi) \bigg)\: p_f(\chi) n^{(3)}_f(\bm{\hat{n}}, \chi).
    \end{align}
    Using the expressions for overdensities Eqs 5, 6,
    \begin{align}
        \frac{n_f(D)}{N_f}(1+\tilde{\delta}^{(3)}_f(\bm{\hat{n}}, D)) = \int d\chi \: p_{h}\left(D-\bar{D}(\chi) - \delta D(\bm{\hat{n}}, \chi) \bigg| z(\chi) \right) p_f(\chi)(1+\delta^{(3)}_f(\bm{\hat{n}}, \chi)).
    \end{align}
The mean of this expression, obtained by dropping the perturbative quantities, yields a prediction for the FRB DM distribution:
    \begin{align}\label{eq:nfD}
        \frac{n_f(D)}{N_f} = \int d\chi \: p_{h}\left(D-\bar{D}(\chi)  \big| z(\chi) \right) p_f(\chi).
    \end{align}
    We now make a perturbative substitution to remove the overdensity from within the host DM distribution. By assuming that the redshift evolution of the host distribution is a first-order quantity, this contribution can be ignored. 
    \begin{align}
        \chi' &\equiv \chi + \frac{\delta D(\bm{\hat{n}}, \chi)}{n_{e, 0}(1+z)}
    \end{align}
With this substitution, we obtain
    \begin{align}
        \frac{n_f(D)}{N_f}(1+\delta^{(3)}_f(\bm{\hat{n}}, D)) &= \int d\chi' \left[ 1 + \frac{\delta D(\bm{\hat{n}}, \chi)}{n_{e, 0}(1+z)^2}\frac{dz}{d\chi} - \delta^{(3)}_e(\bm{\hat{n}}, \chi)\right]  \: p_{h}\left(D-\bar{D}(\chi')\bigg| z(\chi') \right) \notag \\ &\qquad \times \left[ p_f(\chi) - \frac{dp_f(\chi)}{d\chi}\frac{\delta D(\bm{\hat{n}}, \chi)}{n_{e, 0}(1+z)} \right] (1+\delta^{(3)}_f(\bm{\hat{n}}, \chi)).
    \end{align}
 Grouping terms by $\delta D, \: \delta_e$ and subtracting off the expectation yields the DM-space FRB overdensity:
    \begin{align}
        \tilde{\delta}^{(3)}_f(\bm{\hat{n}}, D) = \frac{N_f}{n_f(D)}\int d\chi  \: p_f(\chi)\:p_h\left(D-\bar{D}(\chi)\big| z(\chi)\right) \left[ \frac{\delta D(\bm{\hat{n}}, \chi)}{n_{e, 0}(1+z)}\bigg( \frac{1}{1+z}\frac{dz}{d\chi}- \frac{p'_f(\chi)}{p_f(\chi)}\bigg) - \delta^{(3)}_e(\bm{\hat{n}}, \chi) + \delta^{(3)}_f(\bm{\hat{n}}, \chi)\right].
    \end{align}
    To compute the cross-correlation of angular harmonic coefficients with the spatial galaxy field, we introduce
    \begin{align}
        \delta_g^{(3)}(\bm{\hat{n}}, z_g) = \int d\chi \: p_g(\chi|z_g)\:  \delta_g(\bm{\hat{n}}, \chi) \Rightarrow \int d\chi \: \delta_D(\chi-\chi_g) \: \delta_g^{(3)}(\bm{\hat{n}}, \chi)
    \end{align}
    where we assume a thin galaxy redshift distribution for simplicity of expression, but practically the full integral would be computed. We write spherical harmonic coefficients and the cross-correlation as
    \begin{align}
        a_{lm}^{X} &\equiv \int \frac{d\Omega(\bm{\hat{n}})}{4\pi}Y_{lm}^*(\bm{\hat{n}}) \delta_X(\bm{\hat{n}})\\
    C_l^{fg}(D, z_g) &= \langle a_{lm}^f(D) a_{lm}^{g*}(z_g)\rangle\notag \\
        &= \frac{N_f}{\bar{n}_f(D)}\bigg \langle \int \frac{d\Omega(\bm{\hat{n}})}{4\pi}Y_{lm}^*(\bm{\hat{n}}) \int \frac{d\Omega'(\bm{\hat{n}}')}{4\pi}Y_{lm}(\bm{\hat{n}}')\int d\chi \int d\chi' \delta_D(\chi'-\chi_g) \; p_f(\chi)\:p_h\left(D-\bar{D}(\chi)| z(\chi)\right)\notag \\ &\quad \left[ \frac{\delta D(\bm{\hat{n}}, \chi)}{n_{e, 0}(1+z)}\bigg( \frac{1}{1+z}\frac{dz}{d\chi}- \frac{p'_f(\chi)}{p_f(\chi)}\bigg) - \delta_e(\bm{\hat{n}}, \chi) + \delta_f(\bm{\hat{n}}, \chi)\right]\delta_g(\chi', \bm{ \hat{n}}') \bigg \rangle
\end{align}
We now insert the Fourier space representations of each spatial field and the Rayleigh plane wave expansion for the complex exponentials $e^{i \bm{k}\cdot \chi \hat{n}}$ that appear:
\begin{align}
    C_l^{fg}(D, z_g) &= \frac{N_f}{\bar{n}_f(D)}\bigg \langle \int \frac{d\Omega(\bm{\hat{n}})}{4\pi}Y_{lm}^*(\bm{\hat{n}}) \int \frac{d\Omega'(\bm{\hat{n}}')}{4\pi}Y_{lm}(\bm{\hat{n}}')\int d\chi \int d\chi' \delta_D(\chi'-\chi_g) \; p_f(\chi)\:p_h\left(D-\bar{D}(\chi)| z(\chi)\right)\notag\\ 
    &\quad  \int \frac{d^3\bm{k}}{(2 \pi)^3} \;4 \pi \sum_{l', m'} i^{l'} j_{l'}(k \chi) Y_{l'm'}^*(\hat{\bm{k}}) Y_{l' m'}(\hat{\bm{n}}) \bigg( \frac{\delta D(\bm{k}, z( \chi))}{n_{e, 0}(1+z)}\bigg( \frac{1}{1+z}\frac{dz}{d\chi}- \frac{p'_f(\chi)}{p_f(\chi)}\bigg) - \delta_e(\bm{k}, z(\chi)
   + \delta_f(\bm{k}, z(\chi)) \bigg) \notag\\
    &\quad\ \int \frac{d^3\bm{k'}}{(2 \pi)^3} \;4 \pi \sum_{l'', m''} i^{l''} j_{l''}(k' \chi') Y_{l''m''}^*(\hat{\bm{k}}') Y_{l'' m''}(\hat{\bm{n}}') \delta_g(\bm{k}', z(\chi'))\bigg \rangle
\end{align}
First, we express $\delta D(\bm{k}, z( \chi))$ as an integral over $\chi''$ of $\delta_e(\bm{k})$ with the appropriate factors. Then the integrals over $d\Omega(\bm{\hat{n}})$ and $d\Omega'(\bm{\hat{n}}')$ can be performed using the orthogonality of spherical harmonics. The integrals over $d^3\bm{k}$ and $d^3\bm{k'}$ are changed into spherical coordinates, and the integrals over $d\Omega(\bm{\hat{k}})$ and $d\Omega'(\bm{\hat{k}}')$ are performed similarly. Next, the expectation of combinations of $\langle \delta_X(\bm{k}) \delta_g(\bm{k}')\rangle$ are evaluated, yielding the corresponding power spectra and momentum conserving Dirac delta functions. Finally, we use the Limber approximation to take the integral over $k = |\bm{k}|$, i.e. the radial coordinate, which collapses the two spherical Bessel functions into a Dirac delta. Combining these steps yields:
\begin{align}
    C_l^{fg}(D, z_g) &= \frac{N_f}{\bar{n}_f(D)}\int d\chi' \delta_D(
    \chi'-\chi_g) \int d\chi  \: p_f(\chi)\:p_h\left(D-\bar{D}(\chi)|z(\chi)\right)\notag \\ &\quad \bigg[ \frac{1}{n_{e, 0}(1+z)}\bigg( \frac{1}{1+z}\frac{dz}{d\chi}- \frac{p'_f(\chi)}{p_f(\chi)}\bigg) \left(\int_0^\chi d\chi'' n_{e, 0}(1+z'')P_{eg}\left( \frac{l}{\chi''}, \mu_k=0, z(
    \chi'') \right)\frac{\delta_D(\chi''-\chi')}{\chi''^2} \right) \notag \\ &\quad
    + \left(P_{fg}\left(\frac{l}{\chi}, \mu_k=0, z(\chi)\right) - P_{eg}\left( \frac{l}{\chi}, \mu_k=0, z(\chi)\right)\right)\frac{\delta_D(\chi-\chi')}{\chi^2}  \bigg].
\end{align}
Taking the integral over the narrow galaxy distribution we obtain the final result:
\begin{align}\label{eq:apdxA_final}
    C_l^{fg}(D, z_g) &=   \bigg\{ \frac{N_f}{n_f(D)}\int_{\chi_g}^{\infty} d\chi   \: \:p_h\left(D-\bar{D}(\chi)-\bar{D}_h(\chi)|z\right)  \frac{1}{(1+z)}\bigg( \frac{p_f(\chi)}{1+z}\frac{dz}{d\chi}- p'_f(\chi) \bigg)  (1+z_g)P_{eg}\left( \frac{l}{\chi_g}, \mu_k=0, z_g \right)\frac{1}{\chi_g^2} \bigg \} \notag \\ &\quad
    + \bigg \{\frac{N_f}{n_f(D)}  \: p_f(\chi_g)\:p_h\left(D-\bar{D}(\chi_g)-\bar{D}_h(\chi_g)\right|z_g)  \left(P_{fg}\left(\frac{l}{\chi_g}, \mu_k=0, z_g\right) - P_{eg}\left( \frac{l}{\chi_g}, \mu_k=0, z_g\right) \right)\frac{1}{\chi_g^2} \bigg\}.
\end{align}

\section{Moments of $C_{\ell}^{fg}(D, z_g)$}
Here we detail the computation of the $n=0, n=1$ "moments" of the $C_l^{fg}$ statistic. We begin with the $0$-th moment, starting with the definition of the $n=0$ moment over Eq. \eqref{eq:Clfg}, and next taking the integral over $p_h\left(D-\bar{D}(\chi)\right|z(\chi)) dD$. This gives $1$, by the definition of a normalized probability distribution:
    \begin{align}
     M[D^0 C^{fg}_l(D)] &=  \int dD  \frac{n_f(D)}{N_f} \bigg \{ \frac{N_f}{n_f(D)} \int d\chi 
    \, p_h\left(D-\bar{D}(\chi)\right|z(\chi)) \notag \\
&\quad \times \frac{1}{(1+z)} \left( 
    \frac{p_f(\chi)}{1+z} \frac{dz}{d\chi} - p_f'(\chi)
    \right)
    \int_0^\chi d\chi' \, p_g(\chi')(1+z') 
    P_{eg}\left( \frac{l}{\chi'}, \mu_k=0, z(\chi') \right) 
    \frac{1}{\chi'^2}\bigg \} \notag \\
&\quad + \int dD  \frac{n_f(D)}{N_f}\bigg \{ \frac{N_f}{n_f(D)} 
    \int d\chi \, \frac{1}{N_g} \frac{dN_g}{d\chi} \, 
    p_f(\chi) \, p_h\left(D-\bar{D}(\chi)-\bar{D}_h(\chi)\right|z(\chi)) \notag \\
&\quad \times \left( 
    P_{fg}\left( \frac{l}{\chi}, \mu_k=0, z(\chi) \right) 
    - P_{eg}\left( \frac{l}{\chi}, \mu_k=0, z(\chi) \right)
    \right) \frac{1}{\chi^2} \bigg \}\notag \\
   M[D^0 C^{fg}_l(D)]&=   \bigg \{ \int d\chi  \frac{1}{(1+z)} \left( 
    \frac{p_f(\chi)}{1+z} \frac{dz}{d\chi} - p_f'(\chi)
    \right)
    \int_0^\chi d\chi' \, p_g(\chi')(1+z') 
    P_{eg}\left( \frac{l}{\chi'}, \mu_k=0, z(\chi') \right) 
    \frac{1}{\chi'^2}\bigg \} \notag \\
&\quad + \bigg \{ 
    \int d\chi \, \frac{1}{N_g} \frac{dN_g}{d\chi} \, 
    p_f(\chi)  \left( 
    P_{fg}\left( \frac{l}{\chi}, \mu_k=0, z(\chi) \right) 
    - P_{eg}\left( \frac{l}{\chi}, \mu_k=0, z(\chi) \right)
    \right) \frac{1}{\chi^2} \bigg \}\notag  
\end{align}
Finally, considering the term with the derivative of the FRB distance distribution, $p_f'(\chi)$, we integrate by parts, finding large cancellation and yielding the end result:
\begin{align}
M[D^0 C^{fg}_l(D)]&=\int d\chi\, p_f(\chi)\, p_g(\chi)\,
P_{fg}\left( \frac{\ell}{\chi}, z(\chi) \right)\, \frac{1}{\chi^2}
\end{align}
Next, for the $n=1$ moment, we proceed in a similar fashion. We begin with the definition, and take the integral over $p_h\left(D-\bar{D}(\chi)\right|z(\chi)) DdD$. Here, with an extra factor of $D$, the integral pulls out the mean of the distribution.
\begin{align}
     M[D^1 C^{fg}_l(D)] &=  \int dD  \frac{n_f(D)}{N_f} D\bigg \{ \frac{N_f}{n_f(D)} \int d\chi 
    \, p_h\left(D-\bar{D}(\chi)\right|z(\chi)) \notag \\
&\quad \times \frac{1}{(1+z)} \left( 
    \frac{p_f(\chi)}{1+z} \frac{dz}{d\chi} - p_f'(\chi)
    \right)
    \int_0^\chi d\chi' \, p_g(\chi')(1+z') 
    P_{eg}\left( \frac{l}{\chi'}, \mu_k=0, z(\chi') \right) 
    \frac{1}{\chi'^2}\bigg \} \notag \\
&\quad + \int dD  \frac{n_f(D)}{N_f}D\bigg \{ \frac{N_f}{n_f(D)} 
    \int d\chi \, \frac{1}{N_g} \frac{dN_g}{d\chi} \, 
    p_f(\chi) \, p_h\left(D-\bar{D}(\chi)-\mu(\chi)\right|z(\chi)) \notag \\
&\quad \times \left( 
    P_{fg}\left( \frac{l}{\chi}, \mu_k=0, z(\chi) \right) 
    - P_{eg}\left( \frac{l}{\chi}, \mu_k=0, z(\chi) \right)
    \right) \frac{1}{\chi^2} \bigg \}\notag  \\
    M[D^1 C^{fg}_l(D)]&=   \bigg \{ \int d\chi (\bar{D}(\chi) + \bar{D}_h(\chi)) \frac{1}{(1+z)} \left( 
    \frac{p_f(\chi)}{1+z} \frac{dz}{d\chi} - p_f'(\chi)
    \right)
    \int_0^\chi d\chi' \, p_g(\chi')(1+z') 
    P_{eg}\left( \frac{l}{\chi'}, \mu_k=0, z(\chi') \right) 
    \frac{1}{\chi'^2}\bigg \} \notag \\
&\quad + \bigg \{ 
    \int d\chi \, (\bar{D}(\chi) + \bar{D}_h(\chi))  \frac{1}{N_g} \frac{dN_g}{d\chi} \, 
    p_f(\chi)  \left( 
    P_{fg}\left( \frac{l}{\chi}, \mu_k=0, z(\chi) \right) 
    - P_{eg}\left( \frac{l}{\chi}, \mu_k=0, z(\chi) \right)
    \right) \frac{1}{\chi^2} \bigg \}\notag  
\end{align}
Following the same integration by parts on the $p_f'(\chi)$ term, we again find large cancellation, yeilding the end result:
\begin{align}
M[D^1 C^{fg}_l(D)]&= \int d\chi\, p_f(\chi) \int_0^\chi d\chi'\, p_g(\chi')\, n_{e,0}(1 + z(\chi'))\,
P_{eg}\left( \frac{\ell}{\chi'}, z(\chi') \right)\, \frac{1}{\chi'^2} 
\notag  \\ & \quad + \int d\chi\, p_g(\chi)\, p_f(\chi) \left( \bar{D}(\chi) + \bar{D}_h(\chi) \right)
\, P_{fg}\left( \frac{\ell}{\chi}, z(\chi) \right) \frac{1}{\chi^2}
\end{align}

\bibliographystyle{apsrev4-2}
\bibliography{lit}

\end{document}